\documentclass{article}
\usepackage{spconf,amsmath,graphicx}
\usepackage{booktabs}
\usepackage{makecell}
\usepackage[T1]{fontenc}
\usepackage{diagbox}
\usepackage{physics}
\usepackage{float}

\title{ASCA: Less Audio Data is More Insightful}
\name{Xiang Li$^{1,2}$ 
\qquad Junhao Chen$^{1,2}$ 
\qquad Chao Li$^{1,2\ast}$ \thanks{*Corresponding author}
\qquad Hongwu Lv$^{1,2}$ }
\address{$^{1}$ College of Computer Science and Technology, Harbin Engineering University, China \\
$^{2}$ Modeling and Emulation in E-Government National Engineering Laboratory, China \\
}

\begin{document}
%
\maketitle
\begin{abstract}
Audio recognition in specialized areas such as birdsong and submarine acoustics faces challenges in large-scale pre-training due to the limitations in available samples imposed by sampling environments and specificity requirements. While the Transformer model excels in audio recognition, its dependence on vast amounts of data becomes restrictive in resource-limited settings. Addressing this, we introduce the Audio Spectrogram Convolution Attention (ASCA) based on CoAtNet, integrating a Transformer-convolution hybrid architecture, novel network design, and attention techniques, further augmented with data enhancement and regularization strategies. On the BirdCLEF2023 and AudioSet(Balanced), ASCA achieved accuracies of 81.2\% and 35.1\%, respectively, significantly outperforming competing methods. The unique structure of our model enriches output, enabling generalization across various audio detection tasks. Our code can be found at https://github.com/LeeCiang/ASCA.
\end{abstract}
\section{Introduction}
\label{sec:intro}

Audio detection is important in several applications, such as music style recognition \cite{c2}, environmental sound detection \cite{c3, c4}, and instrument classification \cite{c5}. Traditionally, audio classification has relied on manually designed features such as spectral features and rhythmic patterns \cite{c6}, as well as statistically based methods such as Gaussian Mixture Models (GMM) \cite{c7}. However, with the rise of deep learning, end-to-end neural network models have begun to make significant progress in audio classification tasks \cite{c8, c9}. Among these models, Recurrent Neural Networks (RNNs) \cite{c10, c11} and Convolutional Neural Networks (CNNs) \cite{c27} have become standard components for dealing with time-series dependencies in audio data.

Recently, Transformer-based models \cite{c12, c13}, in particular the self-attention mechanism, have shown their potential advantages over traditional RNN models in audio classification. These models are able to capture long-range dependencies without the temporal constraints of RNNs. However, audio data often contains a lot of extraneous noise, while differences in recording equipment have a significant impact on the data, as well as a rich hierarchical structure. While Transformers address the issue of efficiency in iterative model evaluation, they may have poorer generalization capabilities than Convolutional Neural Networks.Transformers are very data-intensive \cite{c29} and often require pre-training on large datasets. The lack of pre-training on such a large scale is very detrimental to its performance.

In this work, we present Audio Spectrogram Convolution Attention (ASCA),It is derived from CoAtNet that solves small-scale image datasets, and we apply it to solve small-scale audio datasets, which is the first more comprehensive experiment using this Coatnet-based architecture in the field of audio processing, and in order to comprehensively evaluate its performance, we not only compare with AST \cite{c15} but also with EfficientNet \cite{c18}, but also an in-depth comparison with the MAST structural model \cite{c17}. ASCA is unique in its highly optimized architecture. On several datasets, such as AudioSet \cite{c20}, BirdClef2023 \cite{c21}, and VGGSound \cite{c22}, ASCA achieves optimal performance on small-scale datasets. 

\section{Related work}
\label{sec:format}
Feature extraction research on audio signal datasets has a long history, especially on small-scale datasets, and the use of convolutional architectures as basebone is often considered fruitful, and as deep learning continues to improve, advanced network architectures have been used for audio classification, including convolutional neural networks \cite{c27}, while due to the great success of the self-attention mechanism in the field of NLP \cite{c12} and in the field of vision \cite{c25} with great success, convolution-attention networks \cite{c31,c32} as well as pure-attention networks \cite{c17} have been more widely used in audio processing. To better capture global context over long distances, researchers have introduced self-attention mechanisms.AST \cite{c15} pre-training using imagenet \cite{c26} outperforms previous techniques in multiple audio classification benchmarks; Swin Transformer \cite{c14} devises a shift-window strategy in an image transformer; and AudioCLIP \cite{c16} achieves ambient sound categorization (ESC) tasks with new state-of-the-art results with an accuracy of 90.07\% on UrbanSound8K and 97.15\% on the ESC-50 dataset; MAST \cite{c17} uses hierarchical representation learning for efficient audio classification.MAST uses one-dimensional (and two-dimensional) pooling operations along the temporal (and frequency domains) in different phases, however, however, they do not explore these methods' performance or the modifications that can be made to them in order to train from scratch and perform well with small amounts of data.The study by Gani \cite{c23} et al. (2022) extensively explored ViT training with low data volumes and achieved success on small datasets. Their work is based on learning self-supervised induction biases from small datasets and fine-tuning these biases as a weight initialization scheme.The study by Lee et al \cite{c24} (2021) explores how ViTs can be modified to learn local induction biases. Instead, we will build on their work by exploring the use of hybrid models for training in low audio data conditions and presenting a motivation for using hybrid models in low data conditions, which is different from the above work.

\section{METHOD}
\label{sec:pagestyle}
Figure 1 shows the architecture of the proposed audio spectrogram transformer (audio coatnet).In the case of small data volume, convolution is better than transformer for feature extraction; while self-attention accepts the global space, and since convolution has very good features for recognizing displaced images, then the input t-second audio waveform into a sequence of 128-dimensional log Mel filter bank (fbank) features computed every 10ms using a 25ms Hamming window. We input the spectrogram in a size of $224 \times 224$.

\begin{figure*}
    \centering
    \includegraphics[width=0.98\textwidth]{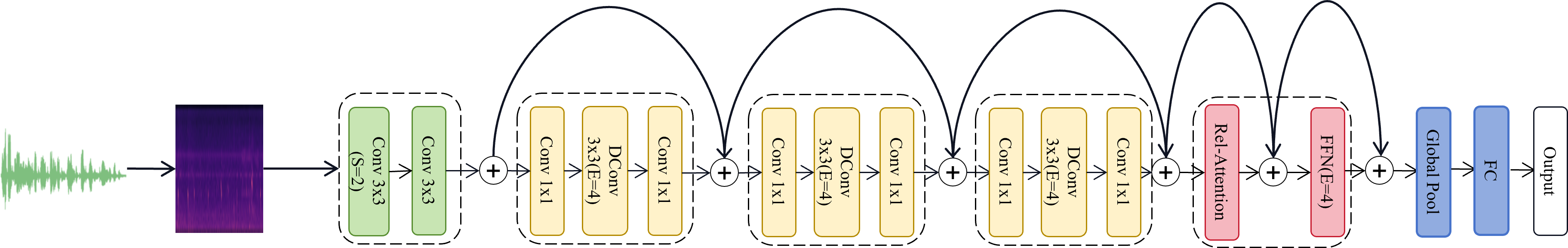}
    \caption{ASCA architecture}
    \label{fig:Descriptive stastistics.}
\end{figure*}

In order to construct a network using relative attention, we adopted an approach similar to CoAtNet, which proposes a network using 5 stages (S0, S1, S2, S3, S4), where S0 is a simple two-layer convolutional starting layer, and S1 uses an inverted residual block with Squeeze-Excitation \cite{c33}. In the study by Dai et al \cite{c1}, they eliminated the possibility of using a C-C-C-T structure because of the allegedly low model performance. However, in our experiments, we found that the C-C-C-T design can achieve better performance in some cases. We believe it is due to this variability caused by the different sizes of datasets, and that the C-C-C-T structure is very well adapted to small datasets. Convolution has the property of translational equivalence. This means that if the input image is translated, the output feature map after convolution will be similarly translated accordingly. The weights of the convolution kernel are pre-learned and are the same for all inputs. Therefore, convolution does not directly support input adaptive weighting. Standard convolution has a localized Receptive Field, which means that each feature of the output is based on a small local region of the input. However, the Receptive Field can be increased indirectly by multilayer convolution and or a large convolution kernel to give it a global characterization. The self-attentive mechanism is not inherently translationally isotropic because it is global and position independent. However, when combined with position coding or other location information, it can realize spatially relevant functions. The attention mechanism assigns weights to each element in the input that are dynamically computed based on the content of the input, thus it implements input adaptive weighting. The self-attention mechanism inherently has a global receiving domain because it considers all positions in the input to compute each position in the output.

In details, the image is first subjected to the operation of convolutional dimensionality reduction, where the model focuses on the MBConv module \cite{c33}. This is an "inverted bottleneck" design, where the input channel size is first expanded by a factor of 4, and then the 4-fold wide hidden state is projected back to the original channel size, and residual joins are then performed between each module. The overall model formulation is as follows, where $x_i$, $y_i$ are the inputs and outputs at location i, respectively, $w_{i-j}$ denotes the deep convolutional kernel, and $\mathcal{G}$ denotes the global space. Here, the attention weights $A_{i, j}$are determined by $w_{i-j}$ and $x_{i}^{\top} x_{j}$. The update of the attention weights is very intuitive and only requires the summation of the global static convolution kernel, the previous convolution module:

\begin{equation}
y_{i}=\sum_{j \in \mathcal{G}} \frac{\exp \left(x_{i}^{\top} x_{j}+w_{i-j}\right)}{\sum_{k \in \mathcal{G}} \exp \left(x_{i}^{\top} x_{k}+w_{i-k}\right)} x_{j}
\end{equation}


It is worth mentioning that, the present model employs a relative self-attention mechanism \cite{c34}, which is a major highlight compared to other audio processing models, where a relative positional encoding is introduced that captures the positional relationship between a query and a key. As a result, the computation of attention is modified as follows: 
\begin{equation}
A_{i, j}=\sum_{k \in \mathcal{G}} \exp \left(x_{i}^{\top} x_{k}\right) \quad \text { (standard self-attention) }
\end{equation}

\begin{equation}
A_{i, j}=\sum_{k \in \mathcal{G}} \exp \left(x_{i}^{\top} x_{k}+\boldsymbol{w}_{\boldsymbol{i}-\boldsymbol{k}}\right) \quad  \text { (relative self-attention) }
\end{equation}

this mechanism has been widely used especially in some variants of the Transformer architecture, such as Transformer-XL, which utilizes the relative attention mechanism to capture a longer range of dependencies, thus improving the performance of long text processing.

\section{Eperiments}
\label{sec:typestyle}

In this section, we focus on evaluating the performance of ASCA on the audio dataset birdCLEF2023.  We will show our results on the main birdCLEF2023 dataset and ablation experiments in Section 4.3 and Section 4.4, respectively. We will then present our experimental results on the ESC-50 dataset, audioset, and Speech Commands V2 dataset in Section 4.3.

\subsection{Dataset and Training Details}
\label{ssec:subhead}
BirdCLEF2023 is made up of short audios of specific birdsongs shared by users of the xenocanto.org platform that constitute the training data. These audios have been tuned to a sampling rate of 32 kHz and converted to ogg format, covering 264 bird species with a total of 16,942 recordings, for which we predicted only the underlying labels.

We also performed related tests on the AudioSet, UCR, and VGG-sound datasets. We took the small AudioSet with a total of 56,246 datasets, but still 527 species. To optimize the data, we used balanced sampling, data amplification and data enhancement techniques (e.g. mixup, masking and background noise).All models are not pre-trained, We used the adamW optimization tool \cite{c35} and the BCE binary cross-entropy loss function to train the model in batches of 12 samples. We tested on standard balanced and complete datasets and evaluated on the BirdCLEF2023 evaluation dataset. During testing, we set a starting learning rate of 5e-5 and performed 30 rounds of training while employing a cosine annealing strategy to adjust the learning rate.

\subsection{Model Optimization}
In the optimization of model performance under low data conditions, we adopt a series of enhancement and regularization strategies. Among them, the enhancement part combines Mixup\cite{c36}, stochastic masking and background noise(0.25)\cite{c37}. As for regularization, we try a variety of strategies, which include random depth regularization, batch normalization, and weighted noise. It is worth noting that batch normalization outperforms other normalization methods for small audio datasets.

\subsection{Results}
\label{sec:Results}
We conducted tests on the BirdCLEF2023, AudioSet(Balanced),
VGG-Sound and UCR datasets to evaluate the performance of various architectures using mAP as a metric for small-scale datasets, and the results are as Table 1.

\begin{table}
\centering
\caption{Experimental Results on Different Datasets}
\label{tab:results}
\begin{tabular}{|c|c|c|c|c|}
\hline
\diagbox[dir=NW]{Model}{Dataset} & \thead{BirdCLEF\\2023\\ (\%)} & \thead{AudioSet\\(balanced)\\ (\%)} & \thead{UCR\\ (\%)} & \thead{VGG-\\Sound\\ (\%)} \\
\hline
Baseline & 77.5 & 25.1 & 85.3 & -  \\
PANNs & 79.1 & 27.5 & 90.9 & -  \\
AST & 80.6 & 33.7 & 93.4 & 78.1  \\
MAST & 80.8 & 34.2 & 93.9 & 81.3  \\
PSLA & 79.2 & 32.7 & 89.2 & 77.8  \\
\hline
\textbf{ASCA(ours)} & \textbf{81.2} & \textbf{35.1} & \textbf{94.2} & \textbf{82.0}  \\
\hline
\end{tabular}
\end{table}

\subsection{Ablation Study}
\label{sec:Ablation Study}
We mainly investigated three experiments, namely the impact of model architecture on performance, the impact of different pre-training data on performance, and the impact of different attention window scales on performance.

\subsubsection{The impact of model architecture}
Our experiments also compared the C-C-T-T architecture and the C-C-C-T architecture, where C is the convolution module and T is the attention module.

\begin{table}[H]
\centering
\caption{Model Architect}
\label{tab:results}
\begin{tabular}{|c|c|}
\hline
\makebox[6em]{Architect} & \thead{result (\%)}\\
\hline
C-C-T-T & 75.3 \\
C-C-C-C & 79.4 \\

\hline
\textbf{C-C-C-T(USED)} & \textbf{81.2} \\
\hline
\end{tabular}
\end{table}
We found that the C-C-C-T architecture can handle small-scale data sets better, and too many self-attention mechanisms cannot It will definitely bring good results. On small-scale data sets, having a suitable combination of convolution and attention is the most important.

\begin{figure*}[htbp]
    \centering
    \includegraphics[width=0.95\textwidth]{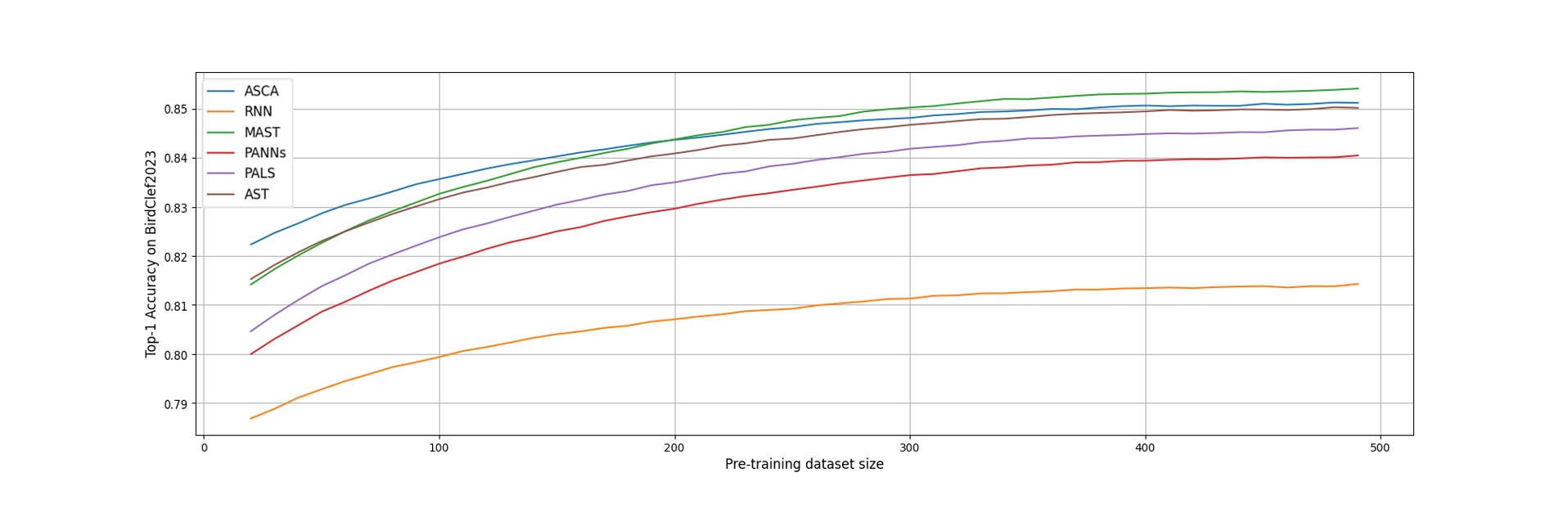}
    \caption{Performance of different models at different pretraining scales.}
    \label{fig:Descriptive stastistics.}
\end{figure*}

\begin{figure}[htbp]
    \centering
    \includegraphics[width=0.45\textwidth]{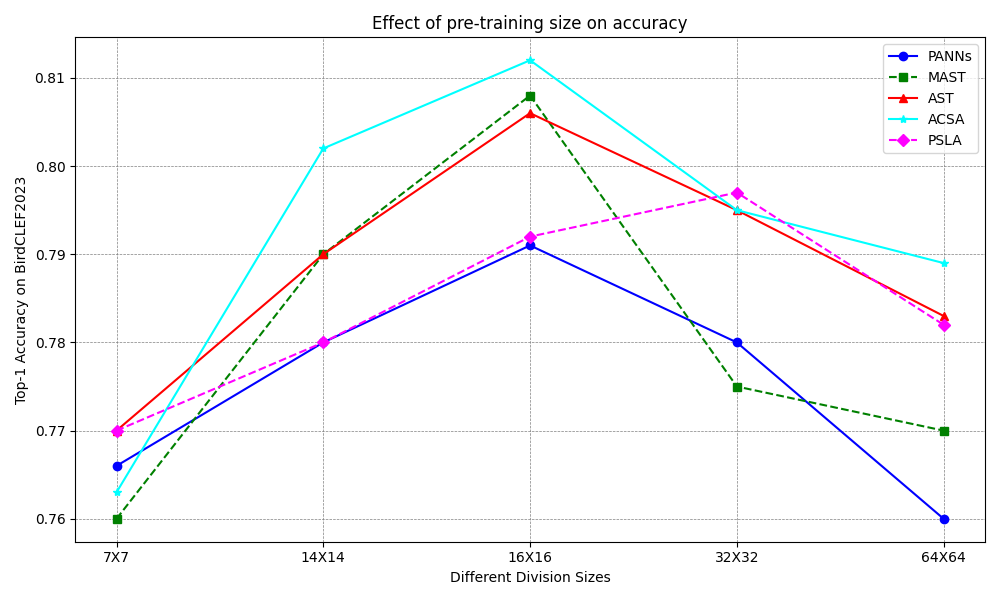}
    \caption{The impact of different partition window sizes on the performance of the attention model}
    \label{fig:Descriptive stastistics.}
\end{figure}

\subsubsection{The impact of different pre-training scales}
\label{sec:Ablation Study}

As can be seen from Figure 2, using the ASCA architecture data set to train the model, we conducted experiments on pre-training data sets of different sizes.on small-scale data sets, the effect is better than all current mainstream models. As shown in Figure 2.

\subsubsection{Multi-scale QKV Spectrogram Feature Extraction}
We investigated the effect of different ViT window divisions on the results, since the input image is 224$\times$224, we divided the following sizes: $7\times7$, $14\times14$, $16\times16$, and $32\times32$, and let's observe the effect of these models with different division sizes.

With different division windows, we find that in many cases 16$\times$16 is by far the best choice, and in the case of 16$\times$16 divisions, ASCA is still the best choice to capture the global information.The results are shown in Figure 3.

\section{Conclusion}
In this work, we presented the evaluation of audio\_CoAtNet on several audio datasets, with a primary focus on the birdCLEF2023 dataset. Our findings revealed that the BirdCLEF2023 dataset, derived from user-shared audios of birdsongs on the xenocanto.org platform, serves as a robust foundation for assessing audio classification models. To optimize the dataset performance, we employed various data enhancement techniques, including mixup, masking, and background noise. With the adamW optimizer\cite{c35} and BCE binary cross-entropy loss function, our training achieved desirable outcomes.

In the journey towards model optimization, we integrated several enhancement and regularization strategies. Notably, batch normalization emerged as a superior approach, especially for smaller audio datasets. Our experiments on various datasets, as documented in Table \ref{tab:results}, highlight the supremacy of our proposed ASCA model. This model not only outperformed others in the BirdCLEF2023 dataset but also showcased exemplary results in other datasets.

In short, the reason why these models perform so well in low-data-mechanism tasks,is, in our opinion, as follows: augmentation and regularization are very important, especially on smaller datasets. The hybrid transformer-convolutional modeling approach is highly generalizable and does not face the problem of highly unstable training.

The ablation study conducted emphasized the significance of pre-training, especially when leveraging the ImageNet dataset.Our results further demonstrate that the ASCA architecture provides a better training foundation for small-scale datasets compared to MAST. In addition, the CCCT architecture proved to be advantageous in handling small-scale datasets. Our exploration into the ViT window divisions for spectrogram feature extraction identified 16$\times$16 as an optimal choice, especially in scenarios where global information capture is paramount. The ASCA architecture still outshone others, emphasizing its ability to process and interpret audio data effectively.

In conclusion, the ASCA model, combined with the appropriate dataset optimization and enhancement techniques, offers promising results in audio classification tasks. Our experiments on various datasets emphasize the model's versatility and effectiveness. The insights drawn from our research can serve as a guidepost for future studies in audio classification and model optimization.


\ninept
\bibliographystyle{IEEEbib}
\bibliography{strings,refs}

\end{document}